\documentclass[%
 superscriptaddress,
 preprint,
 showpacs,
 showkeys,
 preprintnumbers,
  amsmath,amssymb,
  aps,
 pra,
  longbibliography,
  floatfix,
 ]{revtex4-1}
\usepackage{hyperref}

\usepackage{graphicx}
\usepackage{epstopdf}
\usepackage{eepic}
\usepackage[dvipsnames]{xcolor}
\usepackage{braket}
\usepackage{amsmath}
\usepackage{amsthm}
\usepackage{tikz}

\RequirePackage{times}
\RequirePackage{mathptm}

\usepackage{url}
\usepackage{yfonts}

\theoremstyle{plain}

\DeclareMathOperator{\arcosh}{arcosh}

\newcommand{\iprod}[2]{\langle #1 | #2 \rangle}

\newcommand{\oprod}[2]{| #1 \rangle\langle #2 |}

\begin{document}

\title{Value indefinite observables are almost everywhere}

\author{Alastair A. Abbott}
\email{a.abbott@auckland.ac.nz}
\homepage{http://www.cs.auckland.ac.nz/~aabb009}

\affiliation{Department of Computer Science, University of Auckland,
Private Bag 92019, Auckland, New Zealand}
\affiliation{Centre Cavaill\`es, CIRPHLES, \'Ecole Normale Sup\'erieure, 29 rue d'Ulm, 75005 Paris, France}

\author{Cristian S. Calude}
\email{cristian@cs.auckland.ac.nz}
\homepage{http://www.cs.auckland.ac.nz/~cristian}

\affiliation{Department of Computer Science, University of Auckland,
Private Bag 92019, Auckland, New Zealand}

\author{Karl Svozil}
\email{svozil@tuwien.ac.at}
\homepage{http://tph.tuwien.ac.at/~svozil}

\affiliation{Institute for Theoretical Physics,
Vienna  University of Technology,
Wiedner Hauptstrasse 8-10/136,
1040 Vienna,  Austria}

\affiliation{Department of Computer Science, University of Auckland,
Private Bag 92019, Auckland, New Zealand}

\date{\today}

\begin{abstract}
Kochen-Specker  theorems assure the breakdown of certain types of non-contextual hidden variable theories through the non-existence of global, holistic frame functions;
alas they do not allow us to identify where this breakdown occurs, nor the extent of it.
It was recently shown 
[\href {\doibase 10.1103/PhysRevA.86.062109}{Phys. Rev. A \textbf{86}, 062109 (2012)}]
that this breakdown does not occur {\em everywhere;} here we show that it is maximal in that it occurs {\em almost everywhere,} and thus prove that quantum indeterminacy---often referred to as contextuality or value indefiniteness---is a global property as is often assumed.
In contrast to the Kochen-Specker theorem, we only assume the weaker non-contextuality condition that any potential value assignments that may exist are locally non-contextual.
Under this assumption, we prove that once a single arbitrary observable is fixed to occur with certainty, almost (i.e. with Lebesgue measure one) all remaining observables are indeterminate.
\end{abstract}

\pacs{03.67.Lx, 05.40.-a, 03.65.Ta, 03.67.Ac, 03.65.Aa}
\keywords{Kochen-Specker theorem, quantum value indefiniteness, quantum randomness, logical indeterminacy principle, quantum contextuality}
\preprint{CDMTCS preprint nr. 443}
\maketitle

\section{Introduction}
The {\em Kochen-Specker theorem} \cite{specker-60,kochen1}
proves the impossibility of the existence of certain hidden variable theories for quantum mechanics by showing the existence of a finite set of observables $\mathcal{O}$ for which the following two assumptions cannot be simultaneously true for any given individual system:
(P1) every observable in $\mathcal{O}$  has a pre-assigned definite value; and
(P2) the outcomes of measurements of observables are non-contextual.

Non-contextuality means that the outcomes of measurements of observables
are independent of whatever other
co-measurable observables are measured alongside them.
Due to complementarity, the observables in $\mathcal{O}$ cannot all be simultaneously co-measurable, that is, formally, commuting.

The Kochen-Specker theorem does {\em not explicitly identify}
certain particular observables which violate one or both assumptions (P1) and (P2), but only proves their {\em existence}.
This form of the theorem was amply sufficient for its intended scope, primarily to explore the logic of quantum propositions \cite{specker-60}.
The relation between  value indefinite observables, that is, observables which do not have definite values before measurement,
and quantum randomness in \cite{specker-60,kochen1}, requires a more  precise form of the  Kochen-Specker theorem  in which some value indefinite observables can be located (identified). A stronger form of the Kochen-Specker theorem providing this information was proved in \cite{2012-incomput-proofsCJ}.

In this paper we extend these results to show that indeed all observables on a quantum system must be value indefinite except for those corresponding to the contexts compatible with the state preparation.
While it may seem intuitive that quantum indeterminism is  widespread, it does not follow from existing no-go theorems, so it is important that a theoretical grounding be given to this intuition.
This not only helps provide an information theoretic certification of quantum random bits, but also develops our understanding of the origin of quantum indeterminism.

\section{Logical indeterminacy principle}
Pitowsky \cite{pitowsky:218} (also in the subsequent paper \cite{hru-pit-2003} with Hrushovski)
gave a constructive proof of Gleason's theorem in terms of orthogonality graphs which motivated the study of probability distributions on finite sets of rays. In this context he proved a result called ``the logic indeterminacy principle'' which has a striking similarity with the Kochen-Specker theorem and appears as if it could be used to locate value indefiniteness.
However, as we discuss in this section,  this is not the case.

For the sake of appreciating Pitowsky's logical indeterminacy principle, some definitions have to be reviewed.
According to \cite{hru-pit-2003}, a \emph{frame function} on a set $\mathcal{O}\subset \mathbb{R}^n$ of quantum states in a dimension $n\ge 3$ Hilbert space is a function $p:\mathcal{O}\to [0,1]$ such that:
\begin{enumerate}
	\item If $\{\ket{x_1},\dots,\ket{x_n}\}$ is an orthonormal basis, $\sum_{i=1}^n p(\ket{x_i})=1$, and for $\{\ket{x_1},\dots,\ket{x_k}\}$ orthonormal with $k\le n$, $\sum_{i=1}^k p(\ket{x_i})\le 1$.
	\item For all complex $\alpha$ with $|\alpha|=1$ and all $x\in O$, $p(\ket{x})=p(\alpha \ket{x})$.
\end{enumerate}

A {\em Boolean frame function} is a frame function taking only $0,1$ values, i.e.\ for all $\ket{x}\in\mathcal{O}$, $p(\ket{x})\in\{0,1\}$.

Pitowsky's {\em logical indeterminacy principle} \cite{pitowsky:218} states that
{\em for all
states $\ket{a},\ket{b} \in \mathbb{R}^{3}$ with $0<|\langle a|b\rangle| < 1$,
there exists a finite set of
states $\mathcal{O}$ with $\ket{a},\ket{b} \in \mathcal{O}$
such that there is no Boolean frame function $p$ on $\mathcal{O}$ unless $p(\ket{a})=p(\ket{b})=0$.}

A consequence of this principle is that there is no Boolean frame function $p$ on $\mathcal{O}$ such that $p(\ket{a})=1$. From the  {\em logical indeterminacy principle} we can deduce the Kochen-Specker theorem
by identifying each state with the observable projecting onto it, as
a Boolean frame function simply gives a non-contextual,
value definite
yes-no value assignment, so  (P2) is satisfied.

As noted by Hrushovski and Pitowsky \cite{hru-pit-2003},  the  logical indeterminacy principle  is stronger than the Kochen-Specker theorem  because the result is true for arbitrary frame functions which can take any value in the unit interval $[0,1]$, but which are restricted to Boolean values for $\ket{a},\ket{b}$.

In fact, we may be tempted to use the  logical indeterminacy principle  to ``locate'' a
value indefinite observable. Indeed, if we fix $p$ and  choose $\ket{a} \in \mathbb{R}^{3}$ such that
$p(\ket{a})=1$, then, by the  logical indeterminacy principle, for every distinct non-orthogonal unit vector $\ket{b} \in \mathbb{R}^{3}$
it is impossible to have  $p(\ket{b})=1$ and $p(\ket{b})=0$, hence one could be inclined to conclude that the observable projecting onto $\ket{b}$ is value indefinite.
However, such reasoning would be incorrect because if $p(\ket{b})$ were $1$, then the  logical indeterminacy principle merely
concludes that $p$ {\em does not exist};
the same conclusion is obtained if $p(\ket{b})$ were $0$. Hence, in both cases $p$ {\em does not exist},
so it makes no sense to talk about its values, in particular, about $p(\ket{b})$.
(Pointedly stated, from a physical viewpoint, $p(\ket{a})$ as well as $p(\ket{b})$ could take on any of the four combinations of definite values, provided that
(P1) or (P2) is violated for some other observable in $\mathcal{O}$.
Nevertheless, as we shall demonstrate in Section~\ref{sec:main}, using the formalism of \cite{2012-incomput-proofsCJ},
all observables in $\mathcal{O}$ except $\ket{a}$ and those commuting with $\ket{a}$ are indeed provable value indefinite.)
This means that using the  logical indeterminacy principle we get the same global information derived in the  Kochen-Specker theorem,
namely that some observable in $\mathcal{O}$ has to be value indefinite, and no more.
The reason for this limitation is the use of  frame functions, which by definition must be defined everywhere: they can model
``local'' value definiteness, but not  ``local''
value indefiniteness, which, as in the  Kochen-Specker theorem, ``emerges'' only as a global phenomenon.

\section{Value indefiniteness and contextuality}

To remedy the above deficiency we will use the formalism proposed in \cite{2012-incomput-proofsCJ}
for pure quantum states. Specifically, we define value (in)definiteness and contextuality in the framework of quantum logic of Birkhoff and von Neumann \cite{v-neumann-55,birkhoff-36}
and Kochen and Specker \cite{kochen2,kochen3}.

Projection operators projecting on to the linear subspace spanned by a non-zero vector $\ket{\psi}$
will be denoted by  $P_\psi = (\oprod{\psi}{\psi})/ \iprod{\psi}{\psi} $.

Let $\mathcal{O}=\{P_{\psi_1},P_{\psi_2},\dots \}$ be a non-empty set of {\em projection observables} in the $n$-dimensional Hilbert space $\mathbb{R}^n$.
A \emph{context} $C= \{P_1,P_2,\dots P_n\}$ is a set of $n$ orthogonal and thus compatible (i.e.\ simultaneously co-measurable) projection observables from $O$.
In quantum mechanics this means the observables in $C$ are pairwise commuting.
In general, the result of a measurement may depend not just on the observable measured but also on the context it is measured in.
We represent the fact that the measurement of an observable $o$ measured in the context $C$ may be predetermined (e.g. by a hidden variables theory) by a \emph{value assignment function}
which assigns the value $v(o,C)\in \{0,1\}$ to this observable if it is predetermined.
If the result is not predetermined the value $v(o,C)$ is undefined. Formally this means $v$ is in general a partial function.
Accordingly we adopt the convention that $v(o,C)=v(o',C')$ if and only if $v(o,C)$ and $v(o',C')$ are both defined and take equal values.
In what follows, this value assignment function will allow us to formalise the necessary notions of admissibility, value definiteness and non-contextuality.

 To agree with the predictions of quantum mechanics---which place
 certain relations between the values assigned to observables (in any context $C$)---we need to work  with a class of value assignment functions called \emph{admissible}: they are value assignment functions $v$ which satisfies
 the following two properties:
		(i) if there exists an observable $o$ in $C$ with $v(o,C)=1$, then $v(o',C)=0$ for all other observables $o'$ in $C$;
		(ii) if there exists an observable $o$ in $C$ such that for all other observables $o'$ in $C$ $v(o',C)=0$, then $v(o,C)=1$.

	Value definiteness formalises the notion that the result of a measurement (in a particular context) may be predetermined.
For a given value assignment function $v$, an observable $o$ in the context $C$ is {\em value definite} in $C$ if $v(o,C)$ is defined;
otherwise $o$ is \emph{value indefinite} in $C$.
If $o$ is value definite in all contexts then we simply say that $o$ is value definite.

 Non-contextuality corresponds to the classical notion that the value obtained via measurement is independent of other compatible observables measured alongside it.
 An observable $o$ is \emph{non-contextual} if for all contexts $C,C'$ we have $v(o,C)=v(o,C')$; otherwise, $v$ is \emph{contextual}.
The set of observables $\mathcal{O}$ is \emph{non-contextual} if every observable $o\in\mathcal{O}$ is non-contextual;
otherwise, the set of observables $\mathcal{O}$ is \emph{contextual}.
(Here the term \emph{contextual} means that the outcome of a measurement either
exists but is {\em context dependent}, or  it is {\em value indefinite.})

Our definitions of both value definiteness and non-contextually are formulated in a very flexible sense.
They allow us to specify individual value (in)definite observables, and only require observables which are value definite to behave non-contextually.
This technicality is critical in the ability to localise the Kochen-Specker theorem.

\section{Strong Kochen-Specker theorem}

The incompatibility between the assumptions (P1)  and (P2) is not {\em maximal} in the following sense:
for any set of
observables, there exists an admissible assignment function under which the set
of observables is value definite and at least one observable is non-contextual.
This shows that not all observables need to be value indefinite \cite{2012-incomput-proofsCJ},
because for every pure quantum state at least the propositions associated with the state preparation are certain, and thus value definite.

However, {\em there always exist pairs of observables such that, if one of them is assigned the value $1$
by an admissible assignment function under which $\mathcal{O}$ is non-contextual, the other must be
``value indefinite''.}
This result is deduced in \cite{2012-incomput-proofsCJ}
using the weaker assumption that  not all observables are assumed to be value definite,
formally expressed by the admissibility of $v$. In
 particular,
{\em an observable is deduced to be value definite only when the values of other commuting value-definite
observables require it to be so}.

The theorem derived in Ref.~\cite{2012-incomput-proofsCJ}, and henceforth called the
{\em strong Kochen-Specker theorem,}
can be used to ``locate'' a provable value indefinite observable which when measured ``produces'' a quantum random bit,
which is guaranteed to be produced by a value indefinite observable under some physical assumptions:
{\em
       Let $\ket{a}, \ket{b}\in \mathbb{R}^3$ be unit vectors such that $\sqrt{\frac{5}{14}} \le |\iprod{a}{b}| \le \frac{3}{\sqrt{14}}\raisebox{.8mm}{.}$
        Then there exists a set of 24 projection observables $\mathcal{O}$ containing $P_a=\ket{a}\bra{a}$ and $P_b=\ket{b}\bra{b}$ 
		such that there is no admissible assignment function under which $\mathcal{O}$ is non-contextual, $P_a$ has the value $1$ and $P_b$ is value definite.
}

\section{How widespread is value indefiniteness?}
\label{sec:main}

Assuming an observable $P_a$ is predetermined to have the value 1, then
from the strong Kochen-Specker theorem we know that we can explicitly identify an observable $P_b$ which
is provable value indefinite relative to the assumptions (mainly admissibility and non-contextuality).
In this section we address the following question: Which of the remaining observables $P_b$ can be proven to be value indefinite?
We prove here the following answer:
{\em only observables which commute with $P_a$ can be value definite}.

Specifically, we  prove the following, more general  {\em extended Kochen-Specker theorem},
 which  increases the scope of the strong Kochen-Specker theorem to cover the rest of the state space:
{\em
        Let $\ket{a}, \ket{b}\in \mathbb{R}^3$ be neither orthogonal nor parallel unit vectors, i.e. $0 < |\iprod{a}{b}| < 1$.
        Then there exists a set of projection observables $\mathcal{O}$ containing $P_a=\ket{a}\bra{a}$ and $P_b=\ket{b}\bra{b}$
		such that there is no admissible assignment function under which $\mathcal{O}$ is non-contextual, $P_a$ has the value 1 and $P_b$ is value definite. The set $\mathcal{O}$ is finite and can be effectively constructed.
}

While this result is similar in form to the original Kochen-Specker theorem, the subtle differences are critical.
As mentioned previously, the Kochen-Specker theorem is unable to locate value definiteness.
Because if $P_a$ has the value 1,
we cannot conclude that $P_b$ is value indefinite,
even if we can show that any two-valued assignment
leads to a complete contradiction.
This is due to the fact that this contradiction implies only that no \emph{global} assignment function can exist; the Kochen-Specker theorem does not show that $P_b$ could not be value definite, while some other $P_c$ harbours the (necessary) value indefiniteness.

On the other hand, the sets of observables given in the proofs of the stronger form of the
Kochen-Specker theorem presented here
are carefully constructed such that any attempt to place the value indefiniteness on a $P_c$ necessarily contradicts the admissibility of $v$.
For example, it would require a context containing an observable assigned the value 1, and another observable being value indefinite.
This contradicts both the admissibility of $v$, and the physical understanding of what it means for that observable to be assigned the value 1---since we know measuring that observable will give the value 1, measuring the other observables \emph{must} give the value 0, and hence the other observables are necessarily value definite.
As a result, we are forced to conclude that $P_b$ itself is value indefinite.

In order to prove the strong Kochen-Specker theorem,
in Ref.~\cite{2012-incomput-proofsCJ}
a specific proof for the case  $|\iprod{a}{b}|=\frac{3}{\sqrt{14}}$ was given,
followed by a reduction to this proof for the case $|\iprod{a}{b}|<\frac{3}{\sqrt{14}}.$
Here we prove that this theorem can be extended for all cases by reducing
the remaining case of $|\iprod{a}{b}|>\frac{3}{\sqrt{14}}$ to the existing result.
This reduction is more subtle and difficult than the first one.

For the purpose of illustrating the reduction technique, let us state the following {\em reduction lemma}
(derived in Ref.~\cite{2012-incomput-proofsCJ}),
which will also turn out to be important for the reduction we will present later:
{\em Given any two unit vectors $\ket{a},\ket{b}$ with $0 < |\iprod{a}{b}| < 1$ and an $x$ such that $|\iprod{a}{b}| < |x| < 1$, there exist a unit vector $\ket{c}$ with $
\iprod{a}{c}=x$, a set of observables $\mathcal{O}$ containing $P_a=\ket{a}\bra{a}$, $P_b=\ket{b}\bra{b}$, $P_c=\ket{c}\bra{c}$
such that if $P_a$ and $P_b$ have the value 1, then $P_c$ also has the value 1 under any admissible, non-contextual assignment function on $\mathcal{O}$.
Furthermore, if we choose our basis such that $\ket{a}\equiv (1,0,0)$ and $\ket{b}\equiv(p,q,0)$, where $p=\iprod{a}{b}$ and $q=\sqrt{1-p^2}$, then $\ket{c}$ has the form $\ket{c}=(x,y,\pm z)$, where $x=\iprod{a}{c}$, $y=p(1-x^2)/qx$ and $z=\sqrt{1-x^2-y^2}$.}

\begin{figure}[t]
\begin{center}
	\includegraphics{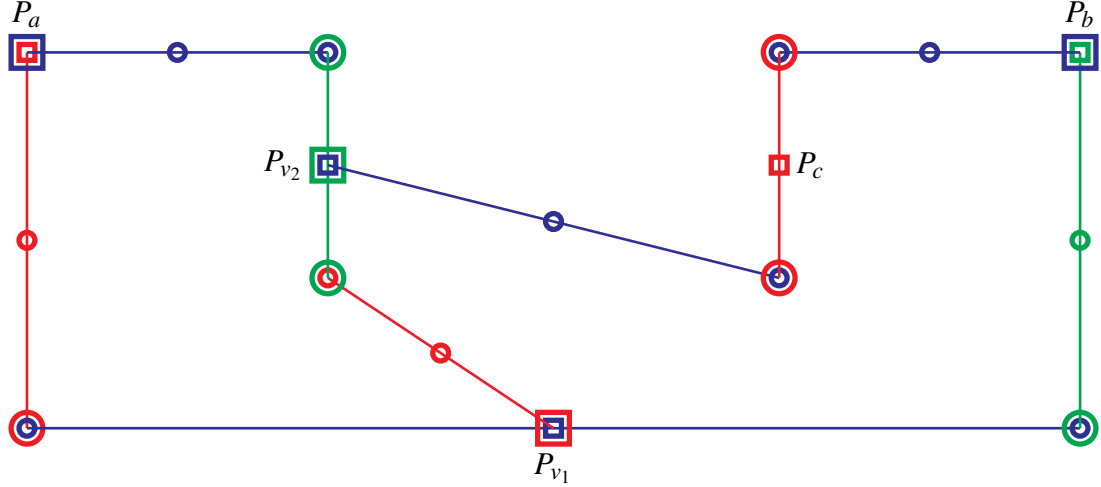}
\end{center}
\caption{(Color online) Greechie orthogonality diagram with an overlaid value assignment that illustrates the reduction in the  iterated reduction lemma.
		}
\label{fig:reduction2}
\end{figure}

This lemma is illustrated in Fig.~\ref{fig:reduction2} and constitutes a simple ``forcing'' of value definiteness: given $P_a$ and $P_b$ both with the value 1, there is a $P_c$ which is ``closer'' (i.e. at a smaller angle of our choosing) to $P_a$ which forces $P_c$ to also take the value 1.

\begin{figure}[t]
\begin{center}
	\includegraphics{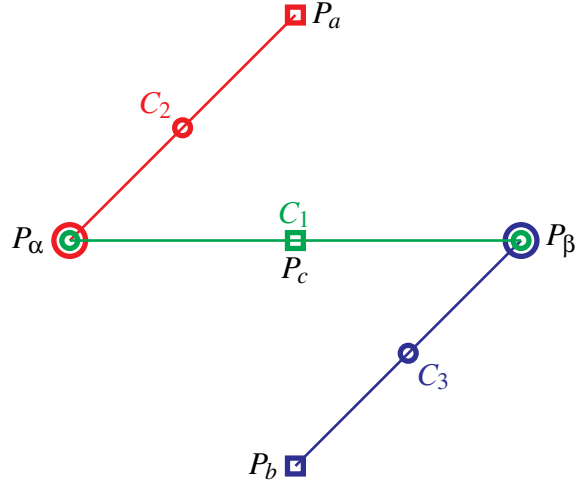}
\end{center}
\caption{(Color online) Greechie orthogonality diagram with an overlaid value assignment that illustrates the reduction in the  reduction lemma.
         The circles and squares represent observables that will be given the values $0$ and $1$ respectively.
         They are joined by smooth lines which represent contexts.}
\label{fig:reduction1}
\end{figure}

This reduction, however, requires necessarily that $|x|>|p|$, and finding a reduction to ``force'' in the other direction (i.e. towards larger angles between $P_a$ and $P_c$) is difficult.
Here we present an argument for this case in what henceforth will be called the {\em iterated reduction lemma}:
{\em
	Given any two unit vectors $\ket{a},\ket{b}$ with $\frac{3}{\sqrt{14}} < \iprod{a}{b} < 1$, there exist
	a unit vector $\ket{c}$ with $\iprod{a}{c}\le\frac{3}{\sqrt{14}}$,
	 a set of observables $\mathcal{O}$ containing $P_a=\ket{a}\bra{a}$, $P_b=\ket{b}\bra{b}$, $P_c=\ket{c}\bra{c}$
	  such that if $P_a$ and $P_b$ have the value 1, then $P_c$ also has the value 1 under any admissible, non-contextual assignment function on $\mathcal{O}$.
}

The proof of this lemma is based on the generalisation of a specific reduction for the case of $\iprod{a}{b}=\frac{1}{\sqrt{2}}$ to $\iprod{a}{c}=\frac{1}{\sqrt{3}}$; that is, it is a ``forcing'' argument in the required direction. The Greechie diagram for this is depicted in Fig.~\ref{fig:reduction2}.
In essence, this figure consists of three copies of the reduction shown in Fig.~\ref{fig:reduction1} glued together, ensuring that the Greechie diagram is indeed realisable. Specifically, the important relations are: $\iprod{a}{v_1}=\sqrt{\frac{2}{3}}\raisebox{0.5ex}{,}$ $\iprod{a}{v_2}=\frac{2}{\sqrt{5}}\raisebox{0.5ex}{,}$ $\iprod{b}{c}=\sqrt{\frac{2}{3}}$ and $\iprod{b}{v_2}=\sqrt{\frac{2}{5}}\raisebox{0.5ex}{.}$
The angles between unit vectors in this proof are then scaled, in a way which we will soon make precise, to fit the required $\iprod{a}{b}$ for the general case.
However, since this doesn't allow us to assert that an arbitrary $\ket{c}$ must have the value 1 in the same way we could using the  reduction lemma, this reduction is then iterated a finite number of times until a sufficiently small $\iprod{a}{c}$ is obtained.

Let us now formally prove the   iterated reduction lemma.
	The constants which will be used for scaling, obtained from the reduction shown in Fig.~\ref{fig:reduction2}, are as follows:
	$$
	\alpha_1=\frac{\arccos\sqrt{\frac{2}{3}}}{\arccos\frac{1}{\sqrt{2}}}\raisebox{0.5ex}{,}\phantom{xxx}
	\alpha_2=\frac{\arccos\frac{2}{\sqrt{5}}}{\arccos\sqrt{\frac{2}{3}}}\raisebox{0.5ex}{,}\phantom{xxx}
	\alpha_3=\frac{\arccos\sqrt{\frac{2}{3}}}{\arccos\sqrt{\frac{2}{5}}}\raisebox{0.5ex}{.}
$$
	Given the initial $\ket{a}$, $\ket{b}$ and the above constants, we thus make use of the following scaled angles between the relevant observables:
	$$
	\theta_{a,b}=\arccos\iprod{a}{b},\phantom{xxx}
	\theta_{a,v_1}=\alpha_1 \theta_{a,b},\phantom{xxx}
	\theta_{a,v_2}=\alpha_2 \theta_{a,v_1}.
	$$
	Once $\ket{v_2}$ is determined via the procedure to follow, we take the following:
	$$
	\theta_{b,v_2}=\arccos\iprod{b}{v_2},\phantom{xxx}
	\theta_{b,c}=\alpha_3 \theta_{b,v_2}.
	$$

	Without loss of generality, let $\ket{a}=(1,0,0)$ and $\ket{b}=(p_1,q_1,0)$ where $p_1=\iprod{a}{b}$ and $q_1=\sqrt{1-p_1^2}$.
	This fixes our basis for the rest of the reduction.
	We want to have $\ket{v_1}$ such that $\iprod{a}{v_1}=x_1=\cos\theta_{a,v_1}$.
	From the   reduction lemma we know this is possible since $x_1>p_1$ (because $\alpha_1 < 1$),
	and we have $\ket{v_1}=(x_1,y_1,z_1)$, $y_1=p_1(1-x_1^2)/q_1 x_1$ and $z_1=\sqrt{1-x_1^2-y_1^2}$.

	We now want $\ket{v_2}$ such that $\iprod{a}{v_2}=x_2=\cos\theta_{a,v_2}$ (this is possible since $\alpha_2 < 1$).
	In order to use the same general form (specified in the reduction lemma) as above, we perform a change of basis to bring $\ket{v_1}$ into the $xy$-plane, describe $\ket{v_2}$ in this basis using the above result, then perform the inverse change of basis.
	Our new basis vectors are given by $\ket{e_2}=(1,0,0)$, $$\ket{f_2}=(\ket{v_1}-x_1\ket{e_2})/q_2=(0,y_1/q_2,z_1/q_2)$$ where $q_2=\sqrt{1-x_1^2}$, and
	$\ket{g_2}=\ket{e_2}\times\ket{f_2}=(0,z_1/q_2,-y_1/q_2)$.
	We thus have the transformation matrix
	$$T_2=
	\begin{pmatrix}
		1 & 0 & 0\\
		0 & y_1/q_2 & z_1/q_2\\
		0 & z_1/q_2 & -y_1/q_2
	\end{pmatrix}\raisebox{0.5ex}{.}
	$$
	\smallskip
	
	We can now put $y_2=x_1(1-x_2^2)/q_2 x_2$ and $z_2=\sqrt{1-x_2^2-y_2^2}$ so that in our original basis we have
	$$\ket{v_2}=T_1 (x_2,y_2,-z_2)^t=(x_2,\frac{y_1 y_2 - z_1 z_2}{q_2}, \frac{y_2 z_1 + y_1 z_2}{q_2}.$$

	We note at this point that the constant $\theta_{b,v_2}$ is now determined, and we have $$\iprod{b}{v_2}=p_1 x_2 + \frac{q_1}{q_2}(y_1 y_2 - z_1 z_2).$$

	For the last iteration of the reduction, we want to find $\ket{c}$ such that $\iprod{b}{c}=x_3=\cos\theta_{b,c}$ (again this will be possible since $\alpha_3<1$).
	Let $p_3=\iprod{b}{v_2}$ and $q_3=\sqrt{1-p_3^2}$.
	Again we perform a basis transformation; we have $\ket{e_3}=\ket{b}=(p_1,q_1,0)$,
	\begin{align*}
		\ket{f_3}&=(\ket{v_2}-p_3\ket{b})/k\\
		&=(x_2-p_3p_1, (y_1y_2-z_1z_2)/q_2-p_3q_1,(y_2z_1+y_1z_2)/q_2)/k,
	\end{align*}
	where $k$ is a constant such that $\ket{f_3}$ is normalized, and
	\begin{align*}
		\ket{g_3}&=\ket{e_3}\times\ket{f_3}\\
		&=\left(\frac{q_1}{q_2}(y_2z_1+y_1z_2),\frac{-p_1}{q_2}(y_2z_1+y_1z_2),\frac{p_1}{q_2}(y_1y_2-z_1z_2)-q_1 x_2\right)/k.
	\end{align*}
	The transformation matrix is then given by
	$$T_3=
	\begin{pmatrix}
		p_1 & \frac{x_2-p_3 p_1}{k} & \frac{q_1(y_2 z_1 + y_1 z_2)}{q_2 k}\\
		q_1 & \frac{y_1 y_2 - z_1 z_2-p_3 q_1 q_2}{q_2 k}& \frac{-p_1(y_2 z_1+y_1 z_2)}{q_2 k}\\
		0 & \frac{y_2 z_1 + y_1 z_2}{q_2 k} & \frac{p_1(y_1 y_2-z_1 z_2)-x_2 q_1 q_2}{q_2 k}
	\end{pmatrix}
	\raisebox{0.5ex}{.}
	$$
	We now put $y_3=p_3(1-x_3^2)/q_3 x_3$ and $z_3=\sqrt{1-x_3^2-y_3^2}$ so that in the original basis we have
	\begin{align*}
		\ket{c}=&T_3 (x_3,y_3,- z_3)^t\\
		=&\left( x_3p_1 + \frac{y_3}{k}(x_2-p_1p_3)- \frac{q_1 z_3}{k q_2}(y_2z_1+y_1z_2),\right.\\
		 		&x_3q_1+\frac{y_3}{k q_2}(y_1y_2-z_1z_2-p_3q_1q_2) + \frac{z_3 p_1}{k q_2}(y_2z_1 + y_1z_2),\\
				&\left.\frac{y_3}{k q_2}(y_2z_1+y_1z_2)- \frac{z_3}{k}\left[\frac{p_1}{q_2}(y_1y_2-z_1z_2)-q_1 x_2\right]\right).
	\end{align*}

	Note that only the first term is of importance in the above expression. Specifically, we want to prove that $\iprod{a}{c}<\iprod{a}{b}=p_1$,
	where
	$$\iprod{a}{c}=x_3p_1 + \frac{y_3}{k}(x_2-p_1p_3)- \frac{q_1 z_3}{k q_2}(y_2z_1+y_1z_2).$$

	The product $\iprod{a}{c}$ is, with appropriate substitutions, a function of one variable, $p_1$; let us denote $f(p_1)=\iprod{a}{c}$.
	We thus need to determine if, for $p_1\in \left(\frac{3}{\sqrt{14}},1\right)$ the inequality $f(p_1)<p_1$ holds.

	We note that $f(p_{1})$ is well behaved and continuous on this domain, and $\lim_{p_1\to 1^-}f(p_1)=1$, hence
	using a combination of direct analysis and symbolic calculation
\cite{Mathematica} and plots, we show that the inequality is indeed true.
	Further details of the analysis are given in the Appendix.

	\begin{figure}
		\begin{centering}
			\includegraphics[scale=0.6]{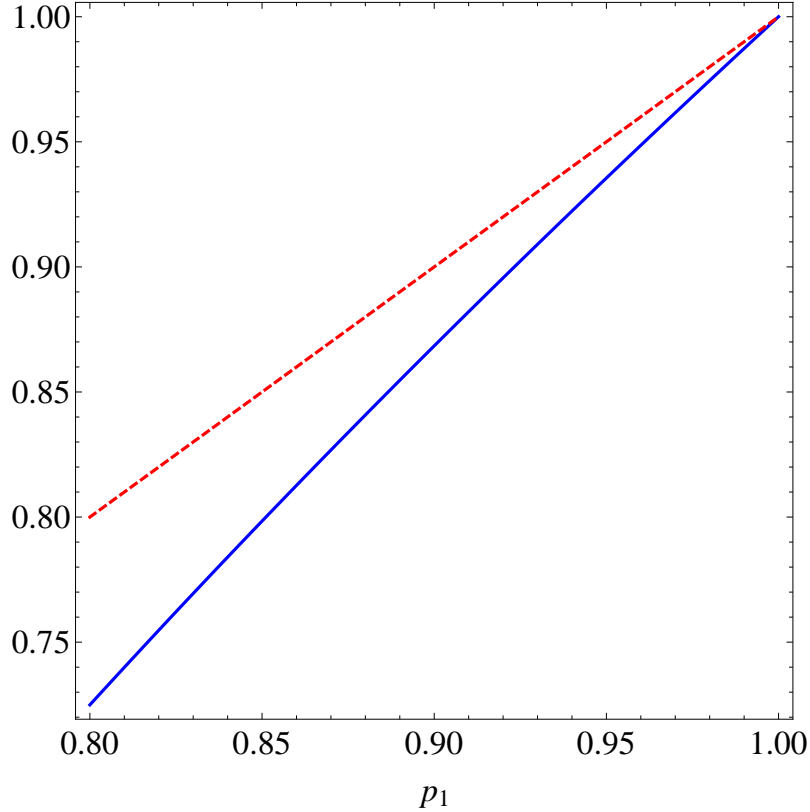}
		\end{centering}
		\caption{(Color online) Plot of $p_1$ (dashed red line) and $f(p_1)$ (solid blue line) for $p_1\in (0.8,1) \supset \left(\frac{3}{\sqrt{14}},1\right)$.}
		\label{fig:mathematicaPlot}
	\end{figure}
	
	Using symbolic calculation
\cite{Mathematica} for a Taylor series expansion around $p_1=1$, we find that for small $|p_1-1|$, $f(p_1)=1-m(1-p_1)$, where $m\approx 1.27$ is a constant. Hence $\lim_{p_1\to 1^-}f(p_1)=1$ as claimed and for some $\varepsilon>0$ we have $f(p_1) < p_1$ for $p_1\in (1-\varepsilon,1)$.
	Further, the continuity of $f$ on this domain can be guaranteed by noting that $f(p_1)$ is simply composed of trigonometric functions with arguments from $(-1,1)\setminus \{0\}$; since these are all continuous, so is $f$.
From Fig.~\ref{fig:mathematicaPlot} and the above results it follows that to prove the inequality   $f(p_1)<p_1$ for all $p_1\in \left(\frac{3}{\sqrt{14}},1\right)$ we need to show that for no $p_1\to 1$
(which implies $f(p_1) \to p_1$) we have $f(p_1)>p_1$.

\begin{figure}
		\begin{centering}
			\includegraphics[scale=0.6]{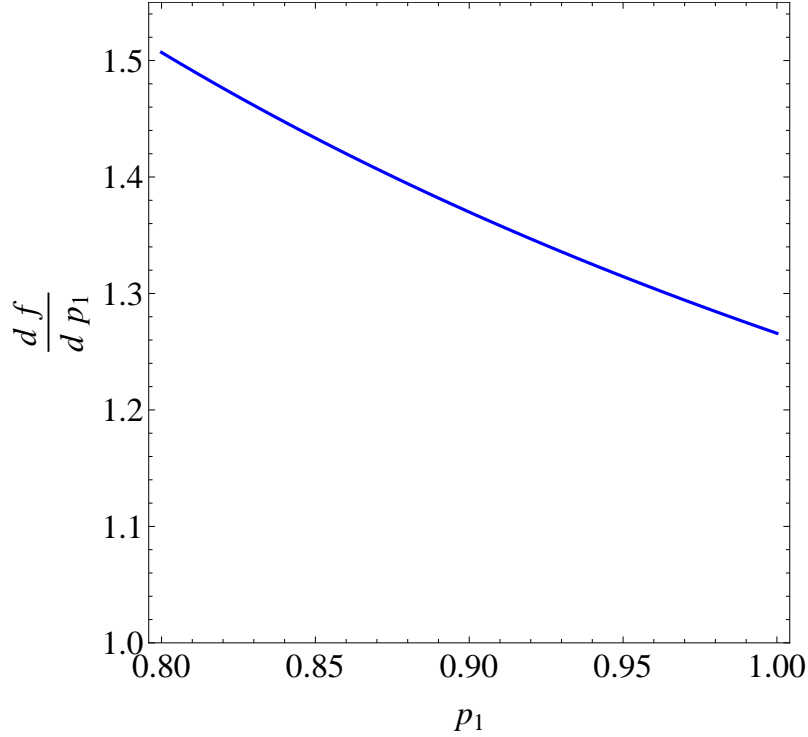}
		\end{centering}
		\caption{(Color online) Plot of $\frac{df}{dp_1}$ for $p_1\in (0.8,1) \supset (\frac{3}{\sqrt{14}},1)$.}
		\label{fig:mathematicaDeriv}
	\end{figure}

	Since  we know from the Taylor series expansion that $f(p_1)<p_1$ in the neighbourhood of $p_1=1$, if for some $p_1'\in \left(\frac{3}{\sqrt{14}},1\right)$ we were to have $f(p_1')>p_1'$, then for some $p_1''$ we must have
	$\frac{d}{dp_1}f(p_1'')<1$, which is false (see Fig.~\ref{fig:mathematicaDeriv}).

	From Fig.~\ref{fig:mathematicaPlot} (and also from the fact that the derivative of $f(p_1) > 1$) it also follows that the difference $p_1-f(p_1)$ is strictly decreasing with $p_1$ on $\left(\frac{3}{\sqrt{14}},1\right) \subset \left(0.8,1\right)$.
	Thus, for large enough (but finite) $k$, $f^k(p_1)\le \frac{3}{\sqrt{14}}\raisebox{0.5ex}{,}$ and the projector $P_{c_k}$ must be assigned the value 1 by $v$.
This completes the proof.

The proof of the  extended  Kochen-Specker theorem  follows rather straightforwardly
from the iterated reduction lemma as follows.
	If $0 < |\iprod{a}{b}| < \frac{3}{\sqrt{14}}$, we can appeal simply to the strong Kochen-Specker theorem, so let $\frac{3}{\sqrt{14}} < |\iprod{a}{b}| < 1$.
	
	Without loss of generality, we can assume that $\iprod{a}{b}\in (0,1)$, since $P_b=P_{\alpha b}$ for $\alpha\in\mathbb{R}$ with $|\alpha|=1$, so the set of projection observables $\mathcal{O}$ obtained under this assumption will give the required result for the general case.
	
	Let us assume, for the sake of contradiction, that such an admissible assignment function $v$ exists for all sets of observables $\mathcal{O}$, i.e. $v(P_a,C_a)=1$ and $v(P_b,C_b)$ is defined for all $C_a,C_b$ with $P_a\in C_a$ and $P_b\in C_a$.
	(Since $v$ is required to be non-contextual, we will omit the context and write $v(P_a,\cdot)$ for simplicity.)
	Then, for all such contexts,
	if $v(P_b,\cdot)=1$, then by the  iterated reduction lemma, there exists a $\ket{c}$ with $\iprod{a}{c}\le\frac{3}{\sqrt{14}}$ such that $v(P_c,\cdot)=1$.
	But this contradicts the   strong Kochen-Specker theorem.
	Hence, if $P_b$ is to be value definite we must have $v(P_b,\cdot)=0$.
	However, we show that this also leads to a contradiction as follows.

    Let $p = \iprod{a}{b}$ and $q = \sqrt{1 - p^2}$.
    We construct an orthonormal basis in which $\ket{a} \equiv (1,0,0)$ and $\ket{b} \equiv (p,q,0)$.
    Define $\ket{\alpha} \equiv (0,1,0)$, $\ket{\beta} \equiv (0,0,1)$ and $\ket{c} \equiv (q,-p,0)$.
    Then $\{ \ket{a}, \ket{\alpha}, \ket{\beta} \}$ and $\{ \ket{b}, \ket{c}, \ket{\beta} \}$ are orthonormal bases for $\mathbb{R}^3$, so we can define the contexts $C_1 = \{ P_a, P_\alpha, P_\beta \}$ and $C_2 = \{ P_b, P_c, P_\beta \}$.
    Since $v(P_a,C_1)=1$, we must have $v(P_\beta,C_1)=v(P_\beta,C_2)=0$ by the admissibility of $v$.
	But since, by assumption, $v(P_b,C_2)=0$, we must have $v(P_c,C_2)=1$.
	However, this also  contradicts the strong Kochen-Specker theorem, since it is easily seen that $$0 < \iprod{a}{c}=q=\sqrt{1-p^2}<\sqrt{\frac{5}{14}}<\frac{3}{\sqrt{14}}\raisebox{0.5ex}{.}$$
	Hence, we conclude that $P_b$ must be value indefinite under $v$.
This then completes the proof the extended Kochen-Specker theorem.

We are now able to answer, in a measure theoretic way,  the question posed in the title of this section:
{\em the set of value indefinite observables  has Lebesgue measure one in $\mathbb{R}^3$.}
The proof starts by noting that the set of value indefinite observables depends on an
arbitrarily fixed single vector, say $\ket{a} \in \mathbb{R}^3$.
Assume that $P_a$ has a definite value (1 or 0). According to the   extended Kochen-Specker theorem,
  no observable outside
the union of the linear subspaces either spanned by the single vector $P_a$ (dimension one),or  the plane orthogonal to this vector $\{P_b \mid \iprod{P_a}{P_b}= 0\}$ (dimension two) is value definite.
This set has Lebesgue measure zero in $\mathbb{R}^3$ because any subset of $\mathbb{R}^3$
 whose dimension is smaller than 3 has Lebesgue measure zero in $\mathbb{R}^3$.
This completes the proof.

In terms of unit vectors,  the set in the above proof corresponds to the set
$\{(1,0,0),(0,0,0)\} \cup \{(0,x,y) \mid x^2 + y^2 = 1\}$ on the three dimensional unit sphere, consisting
of (i) a single point of dimension zero, and (ii) a great circle of dimension one.
Again this set has Lebesgue measure zero on the unit sphere.

\section{Final comments}

One could put our findings in the following perspective.
In response to  Bell-, as well as Kochen-Specker- and Greenberger-Horne-Zeilinger-type theorems,
the ``quantum realists'' -- among them Bell  suggesting that \cite[p.~451]{bell-66}
{\em ``the result of an observation may reasonably depend $\ldots$ on the complete disposition of the apparatus''} --
have been inclined to
adopt {\em contextual value definiteness} in order to save a kind of {\em ``contextual reality.''}
Contextual reality
claims that all measurable properties exist, regardless of whether they are actually measured, or are counterfactuals;
albeit these properties may be context dependent.
In this way one could still maintain the existence of some ``real (though counterfactual, context dependent) physical property.''

While one can probably never rule out such a (necessarily nonlocal) contextual reality,
our results explore the full extent of value indefiniteness.
It is this formalised notion of quantum indeterminism
which can be a crucial element of  quantum information theory,
particularly cryptography and random number generation.

One immediate result of the above findings is that, if one insists on the type of non-contextuality formalized by admissible assignments,
then value definiteness cannot exist outside of a star-shaped configuration in Greechie-type orthogonality diagrams.
It is important to note that this form of non-contextuality is {\em weak}
in the sense that it is only required to apply locally when a definite value is assigned.
Thereby, no holistic frame function on all quantum observables need to be assumed.

Let us be more specific what is meant by the ``star(-shaped)'' configuration of a quantum  state $\vert \psi \rangle$.
We consider a a quantum system prepared in a state corresponding to the proposition that
``a particular detector $D_\psi$ clicks among, say, three
mutual exclusive detectors'' (corresponding to a three dimensional Hilbert space quantum model).
Such a state can be formalized by a projector
$P_\psi = \vert \psi \rangle  \langle \psi \vert$,
or, equivalently, by the linear subspace spanned by the normalized vector $\vert \psi \rangle$
(together maybe with the other two orthonormal vectors to $\vert \psi \rangle$ and to each other).
Now, if a quantum state $\vert \psi \rangle$
is prepared such that the detector $D_\psi$ clicks, that corresponds to  assigning $\vert \psi \rangle$
the value $v( P_\psi ,\cdot )=1$.
$\vert \psi \rangle$'s star is formed by taking some or all vectors $\vert \varphi \rangle$
whose value assignments are consistent
with $v( P_\psi ,\cdot )=1$.
These are value assignments $v( P_\varphi ,\cdot )=0$,
with $\vert \varphi \rangle$ orthogonal to $\vert \psi \rangle$;
that is, $\langle \varphi \vert \psi \rangle=0$.
Such potential observables $\vert \varphi \rangle  \langle \varphi \vert$
are thus value definite.
As they correspond to vectors orthogonal to $\vert \psi \rangle$,
they are, diagrammatically (i.e., in terms of Greechie orthogonality diagrams) speaking,
``in $\vert \psi \rangle$'s star.''

\begin{figure}
\begin{center}¯
\begin{tabular}{ccc}

	\includegraphics{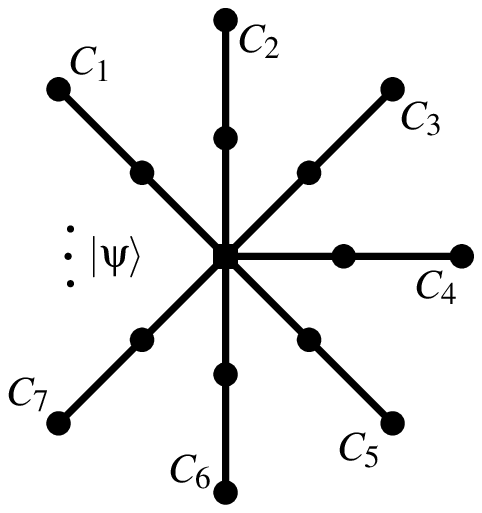}
&
\qquad
\qquad
\qquad
&
	\includegraphics{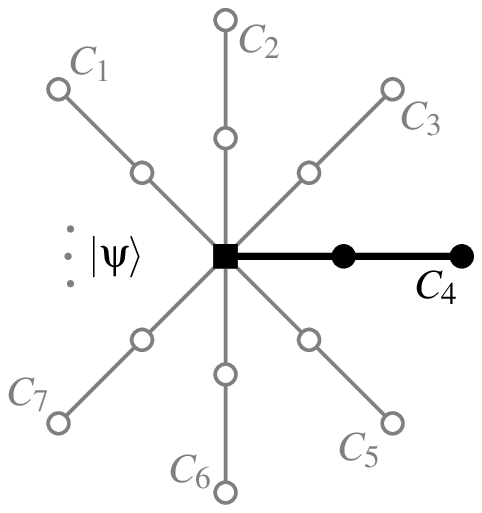}
\\
(a)&&(b)
\end{tabular}
\end{center}
\caption{(Color online)
Greechie orthogonality diagram of a star-shaped configuration,
representing a common detector observable $\vert \psi  \rangle \langle  \psi \vert$
with an overlaid two-valued assignment reflecting $v(P_\psi ,\cdot )=1$.
(a) all ``branches'' corresponding to contexts are assumed to be equally value definite;
(b) it is assumed that, since the system is prepared in, say, context $C_4$, depicted by a block colored in thick filled black,
only this context is value definite;
all the other (continuity of) contexts are   ``phantom contexts'' colored in gray.
}
\label{2013-KstLip-f5}
\end{figure}

All other conceivable observables corresponding to vectors ``outside of $\vert \psi \rangle$'s star''
remain value indefinite relative to our assumptions.
The configuration can be represented by the Greechie orthogonality diagram
depicted in Fig.~\ref{2013-KstLip-f5}(a).
This finding is consistent with the Heisenberg uncertainty relations and quantum complementarity.
Note that this still allows the value definite existence of a continuum of contexts (meaning that all observables therein are value definite) interlinked
at $\vert \psi \rangle$, but on a set of Lebesgue measure zero.

One could be inclined to go one step further and conjecture that
{\em there does not exist any value definite observable outside of a single context} \cite{svozil-2013-omelette}.
This context is defined by the preparation of the state: it consists of
the observable corresponding to $\vert \psi \rangle$,
as well as of the two other orthogonal projectors associated with the two idle detectors that
do not click if $D_\psi $ clicks.
The configuration can be represented by the Greechie orthogonality diagram
depicted in Fig.~\ref{2013-KstLip-f5}(b).
This conjecture is strictly speculative with respect to quantum mechanics, because  with our assumptions
it seems that one cannot prove the  sole existence of just one, unique context among the
continuum of context forming
``$\vert \psi \rangle$'s star.''
Let us mention that one of the authors is inclined to believe in such an existence, another one
is inclined to not believe therein, and the third author has no inclination towards either speculation.

\section*{Acknowledgement} This work was supported in part by Marie Curie FP7-PEOPLE-2010-IRSES Grant RANPHYS.
Part of the work was done while Abbott and Calude were visiting the Institute for Theoretical Physics,
Vienna  University of Technology in September 2013,
and Karl Svozil was visiting the Centre for Discrete Mathematics and Discrete Computer Science,
University of Auckland in February 2014. We thank Michael Trott and Hector Zenil for {\em Mathematica} discussions and help, and Marek Zukowski for critical suggestions which improved the presentation. 
Calude was also supported by a University of Auckland Grant-in-Aid 2013.
Svozil acknowledges Jeffrey Bub, William Demopoulos, and Christopher Fuchs pointing out similarities to Pitowsky's indeterminacy principle.

\appendix*

\section{Further details and code of analysis of $f(p_1)$}
\label{sec:appendix}

The proof of the   iterated reduction lemma   relies critically on the analysis of the function $f(p_1) = \iprod{a}{c}$ for $p_1\in \left(\frac{3}{\sqrt{14}},1\right)$.
Here we give further details of this analysis, which was carried out using Wolfram Mathematica 9.0.1.0.

Specifically, we have
$$f(p_1)=\iprod{a}{c}=x_3p_1 + \frac{y_3}{k}(x_2-p_1p_3)- \frac{q_1 z_3}{k q_2}(y_2z_1+y_1z_2),$$
where the constants are defined in terms of $p_1$ as follows:
$$
\alpha_1=\frac{\arccos\sqrt{\frac{2}{3}}}{\arccos\frac{1}{\sqrt{2}}}\raisebox{0.5ex}{,}\phantom{xxx}
\alpha_2=\frac{\arccos\frac{2}{\sqrt{5}}}{\arccos\sqrt{\frac{2}{3}}}\raisebox{0.5ex}{,}\phantom{xxx}
\alpha_3=\frac{\arccos\sqrt{\frac{2}{3}}}{\arccos\sqrt{\frac{2}{5}}}\raisebox{0.5ex}{,}	$$
$$\theta_{a,b}=\arccos p_1,\phantom{xxx}
\theta_{a,v_1}=\alpha_1 \theta_{a,b},\phantom{xxx}
\theta_{a,v_2}=\alpha_2 \theta_{a,v_1},$$
$$q_1=\sqrt{1-p_1^2},\phantom{xxx}
x_1=\cos\theta_{a,v_1},\phantom{xxx}
y_1=\frac{p_1(1-x_1^2)}{q_1 x_1},\phantom{xxx}
z_1=\sqrt{1-x_1^2-y_1^2},$$
$$q_2=\sqrt{1-x_1^2},\phantom{xxx}
x_2=\cos\theta_{a,v_2},\phantom{xxx}
y2=\frac{x_1(1-x_2^2)}{q_2x_2},\phantom{xxx}
z_2=\sqrt{1-x_2^2-y_2^2},$$
$$p_3=p_1 x_2+q_1\frac{y_1y_2-z_1z_2}{q_2},\phantom{xxx}
\theta_{b,v_2}=\arccos p_3,\phantom{xxx}
\theta_{b,c}=\alpha_3\theta_{b,v_2},$$$$
q_3=\sqrt{1-p_3^2},\phantom{xxx}
x_3=\cos\theta_{b,c},\phantom{xxx}
y_3=p_3\frac{(1-x_3^2)}{q_3 x_3},\phantom{xxx}
z_3=\sqrt{1-x_3^2-y_3^2},$$
$$k=\sqrt{\left(x_2-p_3 p_1 \right)^2 + \left(\frac{(y_1 y_2-z_1 z_2)}{q_2}-p_3 q_1 \right)^2 + \left( \frac{y_2 z_1+y_1 z_2}{q_2} \right)^2}.$$

The Mathematica code used for the analysis (available in \cite{2013-KstLip-RR}) uses these constants and the form of $f(p_1)$ to give the following Taylor expansion of $f$ at $p_1=1$, showing  the behaviour of $f(p_1)$ as $p_1\to 1$ from below.
It also calculates the derivative which is used to generate Fig.~\ref{fig:mathematicaDeriv}.

\begin{align*}
	f(p_1)=&
		1+\frac{(p_1-1)}{\pi ^2 \arccos^2\sqrt{\frac{2}{5}}}
	   \left(\pi ^2 \left(\arccos^2\sqrt{\frac{2}{5}}+
	   \arcosh^2\sqrt{\frac{2}{3}}\right)\right.\\
	   &+\left. 8
	   \arccos\frac{2}{\sqrt{5}} \left(
	   \arccos\frac{2}{\sqrt{5}} \left(2
	   \arccos^2\sqrt{\frac{2}{3}}\right.\right.\right.\\
	   &+\left.\left.\left.\sqrt{\left(\pi ^2+16
	   \arcosh^2\sqrt{\frac{2}{3}}\right) \left(
	   \arccos^2\sqrt{\frac{2}{5}}+
	   \arcosh^2\sqrt{\frac{2}{3}}\right)}\right)+4
	   \arccos\sqrt{\frac{2}{3}}\right.\right.\\
	   &\times\left.\left. \sqrt{\left(
	   \arccos^2\sqrt{\frac{2}{5}}+
	   \arcosh^2\sqrt{\frac{2}{3}}\right) \left(
	   \arccos^2\sqrt{\frac{2}{3}}+
	   \arcosh^2\frac{2}{\sqrt{5}}\right)}\right)\right)\\
	   &+\mathcal{O}((p_1-1)^{2}),
\end{align*}
which numerically simplifies to
$$f(p_1)=1-1.2658(1-p_1)+\mathcal{O}((p_1-1)^2).$$

%

\end{document}